\documentclass[twocolumn,amsmath,amssymb,floatfix,prl]{revtex4}

\usepackage{graphicx}

\def\nm{{\ {\rm nm}}}						
\def\micron{{\ \mu{\rm m}}}					

\def\Hz{{\ {\rm Hz}}}						
\def\kHz{{\ {\rm kHz}}}						
\def\MHz{{\ {\rm MHz}}}						

\def\us{{\ \mu{\rm s}}}						
\def\ms{{\ {\rm ms}}}						

\def\nK{{\ {\rm nK}}}						

\def\Er{{{E_R}}}							
\def\kr{{{k_R}}}							
\def\Rb87{^{87}\rm{Rb}}					
\def\UoverTC{(U/t)_{\rm c}}					

\begin{document}

\title{Condensate fraction in a 2D Bose gas measured across the Mott-insulator transition}

\author{I.~B.~Spielman}
\email{ian.spielman@nist.gov}
\author{W.~D.~Phillips}
\author{J.~V.~Porto}
\affiliation{Joint Quantum Institute, National Institute of Standards and Technology, and University of Maryland, Gaithersburg, Maryland, 20899, USA}

\date{\today}

\begin{abstract}
We realize a single-band 2D Bose-Hubbard system with Rb atoms in an optical lattice and measure the condensate fraction as a function of lattice depth, crossing from the superfluid to the Mott-insulating phase.  We quantitatively identify the location of the superfluid to normal transition by observing when the condensed fraction vanishes.  Our measurement agrees with recent quantum Monte Carlo calculations for a finite-sized 2D system to within experimental uncertainty.
\end{abstract}

\maketitle

Measurements of condensed matter systems realized by cold atoms in optical lattices are now performed with sufficient accuracy to compare with {\it ab-initio} calculations~\cite{Gerbier2005,Spielman2007,Mun2007}.  Bosonic atoms in a lattice nearly perfectly realize the iconic Bose-Hubbard (BH) Hamiltonian.  Here, we study the system's momentum distribution, measure the condensate fraction, and accurately identify the transition point from the the low temperature superfluid (SF) phase by identifying when the condensate fraction vanishes.

The SF to Mott insulator (MI) transition can be accessed by changing the depth of the optical potential~\cite{Jaksch1998}, and has been observed in 1D~\cite{Kohl2005}, 2D~\cite{Spielman2007} and 3D~\cite{Greiner2002}.  A range of studies have verified a detailed understanding of the MI phase in 2D and 3D~\cite{Gerbier2005,Gerbier2005a,Folling2006a,Spielman2007}.  In contrast, the SF phase and the details of the transition to MI have gone largely unstudied.  Indeed, the only quantitative measurement locating the transition is in 3D and is not in agreement with calculations~\cite{Mun2007}.  Here we focus specifically on the superfluid phase of a 2D system and its transition to a normal state: we observe the expected increasing momentum spread and vanishing condensate fraction as the system leaves the SF phase.  Our measured transition point agrees with the best available calculations~\cite{Wessel2004}, thereby locating a point on the non-zero temperature 2D BH phase diagram.  Interestingly, the condensate fraction in our non-zero temperature system vanishes more sharply than expected for a zero temperature inhomogenous system, confirming that the superfluid regions are rapidly driven normal as soon as an insulator appears~\cite{Ho2007,Gerbier2007}.

The physics of interacting systems frequently depends spectacularly on dimensionality: in 3D the SF is a conventional Bose-Einstein condensate (BEC); in 2D, a Berezinskii-Kosterlitz-Thouless (BKT) SF; finally, in 1D there is no true SF.  In contrast, only the detailed properties of the MI phase depend on dimensionality.  In the $T>0$, 2D case studied here, the very existence of Bose condensation is a consequence of the finite size of our trapped system.  We associate the presence of a bimodal momentum distribution with the SF phase, and use fits to the distribution to identify the Bose-condensed fraction and thereby measure the transition point between SF and normal.

At low temperature the transition from SF is to a normal state which crosses over to a MI phase as the lattice depth increases~\cite{Gerbier2007,Ho2007}.  As a result any $T>0$ measurement based on condensate fraction will identify the SF to normal transition but be largely insensitive to the subsequent crossover into the MI phase.

We study samples of ultra-cold rubidium atoms in a combined sinusoidal plus harmonic potential.  For atom occupancy per lattice site larger than unity~\cite{Greiner2002,Campbell2006}, the low temperature SF phase (shallow lattice) is expected to evolve into a structure composed of alternating shells of SF and integer-occupied MI (deep lattice).  As the lattice deepens, each successive MI region appears and grows, as probed in Ref.~\cite{Folling2006a}.  At $T=0$ the amount of SF varies smoothly with lattice depth giving no abrupt changes in the momentum distribution to indicate a phase transition.  In this work, we simplify the situation by working near unit filling, where the only insulating phase is unit occupied MI; thus, any observed signature can only be the transition from SF to normal.  Absent the lattice, recent experiments have shown that weak contact interactions lead to a decrease in the 2D BEC transition temperature~\cite{Kruger2007}.  Lattice potentials increase the relative importance of interactions; indeed, the onset of the MI phase corresponds to driving the critical temperature to zero.

The BH model describes lattice-bosons that have a hopping matrix element $t$, and an on-site interaction energy $U$.  The physics of the BH model depends only on $U/t$~\cite{Fisher1989}.  In an infinite, homogenous $T=0$ 2D system, the transition from SF to MI occurs at the critical value $\UoverTC\approx16.5$~\cite{Krauth1991a,Elstner1999,Wessel2004,Kato2007}.  Remarkably, we observe a sharp transition at $U/t=15.8(20)$~\footnote{All uncertainties herein reflect the uncorrelated combination of single-sigma statistical and systematic uncertainties.} in our $T>0$, finite-sized, harmonically trapped system.

\begin{figure}[t]
\begin{center}
\includegraphics[width= 3.25in]{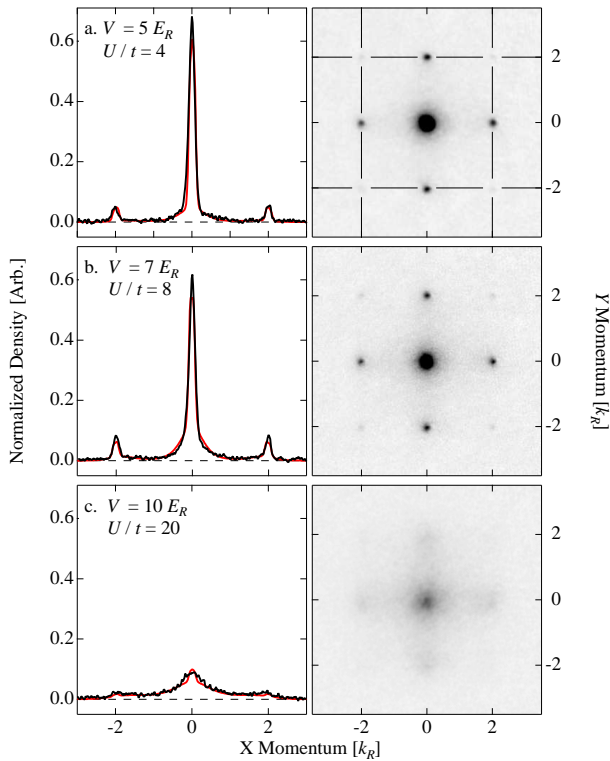}
\end{center}
\caption{Momentum distributions and cross-sections at $U/t$ = 4(1), 8(1), and 20(2).  Each row shows a single momentum distribution normalized by the total atom number; the lines in the top-right panel indicate trajectories along which four cross-sections were taken.  The left panel shows the average of these four sections (black solid line); the red-dashed lines denote the fit to the bimodal distribution.}
\label{rawdata}
\end{figure}

Our data consists of images of atom density after sudden release and time-of-flight (TOF), approximating the {\it in-situ} momentum distribution.  Figure~\ref{rawdata} shows 2D momentum distribution (right), and cross-sections through each distribution (left).  As evidenced by Fig.~\ref{rawdata}-a and -b, each diffraction order in the momentum distributions consists of a narrow peak on a broad pedestal.  Fitting to a bimodal distribution (see below), we determine $f$, the fractional contribution of the narrow component, and identify $f$ as the ``condensate'' fraction.  We associate images with non-negligible $f$ as being in the SF phase~\cite{Yi2007}.  We emphasize, however, that superfluidity is a transport phenomena and cannot unambiguously be associated with features in the momentum distribution~\cite{Diener2007,Gerbier2007,Ho2007}.  This association is also imperfect at $T>0$ because in our 2D trapped system we expect to observe a discernible ``condensate'' fraction even after the vortex-pairs of a BKT SF unbind~\cite{Hadzibabic2006}, destroying the 2D SF.  $f$ vanishes only when the resulting phase-fluctuating quasi-condensate vanishes~\cite{Kruger2007,Clade2008}.

To characterize the transition from SF to normal, we extract two independent quantities from an analysis of TOF images: $f$, and an ``energy scale'' $\sigma$.  We also measure a related quantity, the full width at half maximum (FWHM) $\Gamma$ of the quasi-momentum distribution, which we compare to theory.  As the lattice depth is increased we find that $f$ vanishes concurrently with a sudden increase in $\Gamma$, abrupt signatures that we associate with the transition.

We produce nearly pure 3D $\Rb87$ BECs with $N_T=1.2(4)\times10^5$ atoms in the $\left|F=1,m_F=-1\right>$ state~\cite{Spielman2007}.  A pair of linearly polarized, $\lambda = 820\nm$ laser beams forms a $30(2)\Er$ deep vertical optical lattice along $\hat z$ that divides the 3D BEC into about $70$ 2D systems (turn-on time = $200\ms$).  The single photon recoil wave-vector and energy are $\kr = 2\pi/\lambda$ and $\Er=\hbar^2\kr^2/2 m=h\times3.4\kHz$; $m$ is the atomic mass and $h$ is Planck's constant.  The largest 2D system, containing $\approx3000$ atoms, has a chemical potential $\mu_{\rm 2D} = h\times 600(100)\Hz$ and we measure a temperature $k_B T=k_B\times33(4)\nK = h\times700(70)\Hz$.  Since the first vibrational spacing $h\times33(1)\kHz\gg \mu_{\rm 2D}, \ k_B T$, this system is well into the 2D regime.  In addition, a weaker, square 2D lattice in the $\hat x$-$\hat y$ plane is produced by a second beam arranged in a folded-retroreflected configuration~\cite{Sebby-Strabley2006}, linearly polarized in the $\hat x$-$\hat y$ plane (turn-on-time = $100\ms$~\footnote{The beams for the two lattices originate from the same Titanium-Sapphire laser but differ in frequency by about $160\MHz$.}).  The intensities of both lattices follow exponentially increasing ramps, with $50\ms$ and $25\ms$ time constants respectively, and reach their peak values concurrently.  These time-scales are chosen to be adiabatic with respect to mean-field interactions, vibrational excitations, and tunneling within each 2D system.  The final depth of the $\hat x$-$\hat y$ lattice determines $U/t$ and ranges from $V=0$ to $25(2)\Er$~\footnote{The depth of the lattice along $\hat x$ and $\hat y$ differ by 6\%, and $V$ is the average.}.  The lattice depths are calibrated by pulsing the lattice for $3\us$ and observing the resulting atom diffraction~\cite{Ovchinnikov1998}.

We calculate $U/t$ using a 2D band-stucture model and the $s$-wave scattering length~\cite{vanKempen2002}.  The uncertainty in $U/t$ stems from the uncertainty in lattice depth~\footnote{The uncertainty in the $\hat x$-$\hat y$ lattice depth affects both $t$ and $U$, while vertical lattice uncertainties affect only $U$.  The $\pm 0.2\%$ uncertainty~\cite{vanKempen2002} in the $\Rb87$ $s-$wave scattering length is a negligible contribution to the overall uncertainty.}.  The resulting uncertainty in $U/t$ is $\pm 10\%$.

Once both lattices are at their final intensity, the atomic system consists of an array of 2D gasses each in a square lattice of depth $V$ and with a typical density of 1 atom per lattice site.  The atoms are held for $30\ms$, and all confining potentials are abruptly removed (the lattice and magnetic potentials turn off in $\lesssim1\us$ and $\simeq300\us$, respectively).  As a result, the initially confined states are projected onto free particle states which expand for a $20.1\ms$ TOF~\footnote{Some of the data were not taken under exactly these conditions: some had a $29.1\ms$ TOF and in others the lattice depth was rapidly (in $50\us$) increased to $30\Er$ before imaging (changing the single-site wave functions, not the correlations which structure to the momentum distribution); these differences do not affect the measurement, and are included in the figures.
}, when they are detected by resonant absorption imaging.  Apart from effects of atomic interactions during expansion and the initial size of the sample, initial momentum maps into final position, so each image approximates the $\hat x$-$\hat y$ projection of the momentum distribution.  We fit the each momentum distribution to a simple function which describes the distributions over the full range of $U/t$ studied here, with just three free parameters.

First, we model the broad background as a thermal distribution of non-interacting classical particles in a single 2D sinusoidal band where states are labeled by their quasi-momentum $q_x$ and $q_y$, $n(q_x,q_y) \propto \exp\left[2 (\cos \pi q_x/\kr + \cos \pi q_x/\kr)/\sigma \right]$; this contributes two fitting parameters: $\sigma$ and the non-condensed atom-number.  In the shallow lattice limit, $\sigma$ gives the temperature, $\sigma=k_B T/t$.  This fit does not distinguish atoms thermally occupying higher momentum states from atoms occupying these states in the ground state wavefunction, i.e., from the quantum depletion of the SF.  In fact, $n(q_x,q_y)$ multiplied by a suitable Wannier function, correctly describes the momentum distribution of atoms in the MI phase to first order in $t/U$ where $\sigma$ is unconnected to temperature, and is given by $\sigma = U/4t$.  Our function fits the random phase approximation (RPA) momentum distribution fairly well even as higher order terms become important~\cite{Sengupta2005,Spielman2007}.

The second portion of the momentum distribution consists of a narrow peak, which we interpret as Bose-condensed atoms.  We therefore take the narrow peak to be the inverted parabola of a Thomas-Fermi profile (of fixed width for all comparable data~\footnote{We do not allow the width of the condensate-peak to vary with each fit; instead we first fit all of the SF data with the condensate-width as a free parameter, and then repeat the fits with it held constant at the average value: for $20.1\ms$ TOF we found $R_{\rm TF} = 19(2)\micron$, and for $29.1\ms$ TOF we found $R_{\rm TF} = 26(2)\micron$.}), characterized by a single fitting parameter, condensed number.

The observed condensate peak width after TOF stems largely from initial system size, not interaction effects during TOF or the initial momentum spread.  Here interactions during TOF are reduced due to rapid expansion along $\hat z$ after release from the tightly confining vertical lattice.  Our analysis further reduces these interaction effects by excluding data inside the $1^{\rm st}$ Brillouin zone, with the highest density. This decreases the measured FWHM of the peak from $30(1)\micron$ to $22(1)\micron$ and the inferred momentum width from $0.26\kr$ to $0.21\kr$.  Changing the TOF from $20.1\ms$ to $29.1\ms$ only increased the FWHM from $22(1)\micron$ to $28(1)\micron$ (decreasing the observed momentum width from $0.21\kr$ to $0.17\kr$).

\begin{figure}[t]
\begin{center}
\includegraphics[width= 3.25in]{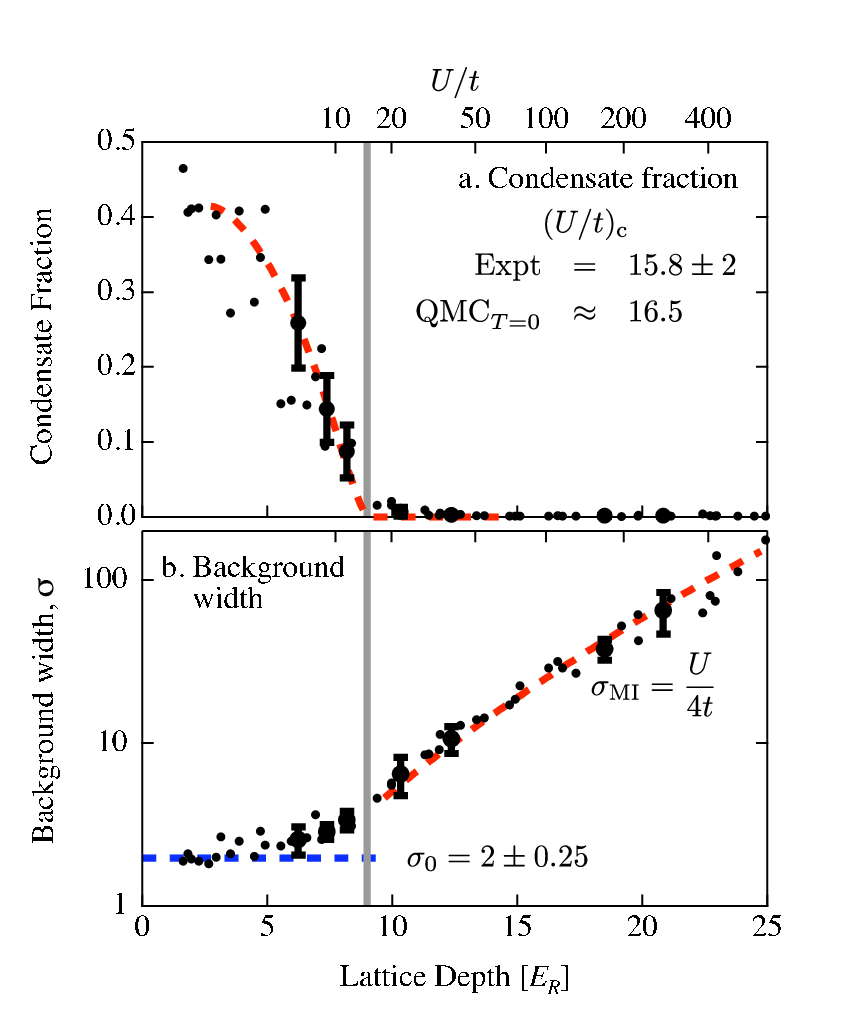}
\end{center}
\caption{Condensed fraction $f$ and $\sigma$ vs. $V$ (bottom axis) or $U/t$ (top axis).  The dots denote values determined from 2D fits to the full momentum distribution: small dots result from one image and the large dots indicate data averaged over about 20 separate images.  The uncertainties are their RMS variation, and are indicative of the single-image uncertainties.  (a) Condensate fraction.  The red-dashed line is computed from our MFT model.  (b) fit parameter $\sigma$. There are two distinct regimes: at low $U/t$ it is nearly constant (blue dashed line), from which we infer an initial temperature $k_B T \approx 2 t$; and at large $U/t$ it monotonically increases, consistent with predictions of perturbation theory in the MI phase (red-dashed line).}
\label{fract}
\end{figure}

Figure~\ref{fract}-a shows that as $V$ increases, $f$ vanishes at a critical value $V_{\rm crit}$, while the total atom number remains constant.  We verified that this disappearance does not result from excessive irreversible heating of the system by exceeding $V_{\rm crit}$, then lowering the lattice and observing a condensed fraction~\cite{Greiner2002}.

To gain a {\it qualitative} understanding of the vanishing condensate fraction, we performed a non-zero temperature mean-field theory (MFT) simulation of an array of 2D BH systems in a 3D harmonic trap~\cite{Sheshadri1993}.  To model the non-zero temperature experimental system, we determine the entropy at small $U/t$ that gives the observed $\approx45\%$ condensate fraction, and assume this entropy is unchanged as $V$ increases.  The red-dashed line in Fig.~\ref{fract}-a shows the MFT condensate fraction vs. $V$ at constant entropy.  Given that $T=0$ MFT overestimates the transition ($(U/t)_{\rm MFT}=23.3$, compared to $\UoverTC=16.5$ from more accurate calculations), the curve unexpectedly lies on the data.  MFT also gives $f$ as function of $U/t$ in units of $\UoverTC$.  We identify the transition point by fitting this function to the data allowing $\UoverTC$ to vary, yielding $\UoverTC=15.8(20)$ (a lattice depth $V_{\rm crit}=9.0(5)\Er$).

Figure~\ref{fract}-b displays $\sigma$ from the uncondensed background portion of the distribution.  At large $V$ we recover the behavior expected in the MI phase; this measurement is equivalent to observations of the modulated momentum distribution in the MI phase~\cite{Gerbier2005,Spielman2007}.  At higher total atom number, our system would develop doubly and triply occupied MI shells, expected to manifest as kinks in this curve.  $\sigma$ is monotonic with $V$, varying smoothly across $V_{\rm crit}$.  This is in agreement with RPA theory where the onset of superfluidity affects only states near zero quasi-momentum.  Figure~\ref{fract}-b shows that when $V\lesssim4\Er$ ($U/t \lesssim 3$), $k_B T/t \approx 2.0(3)$.  Extrapolating to $V=0$ gives $k_B T = \Er\sigma/\pi^2 \approx k_B\times33\nK$ (valid when $T\ll\Er$).  This temperature is well below the $k_B\times45\nK$ expected for non-interacting particles in our 2D harmonic trap with $f=0.45$, this reduction is similar to that observed in Ref.~\cite{Kruger2007}, which focused on the critical temperature in interacting 2D atomic systems with no 2D lattice.

\begin{figure}[tbhp]
\begin{center}
\includegraphics[width=3.25in]{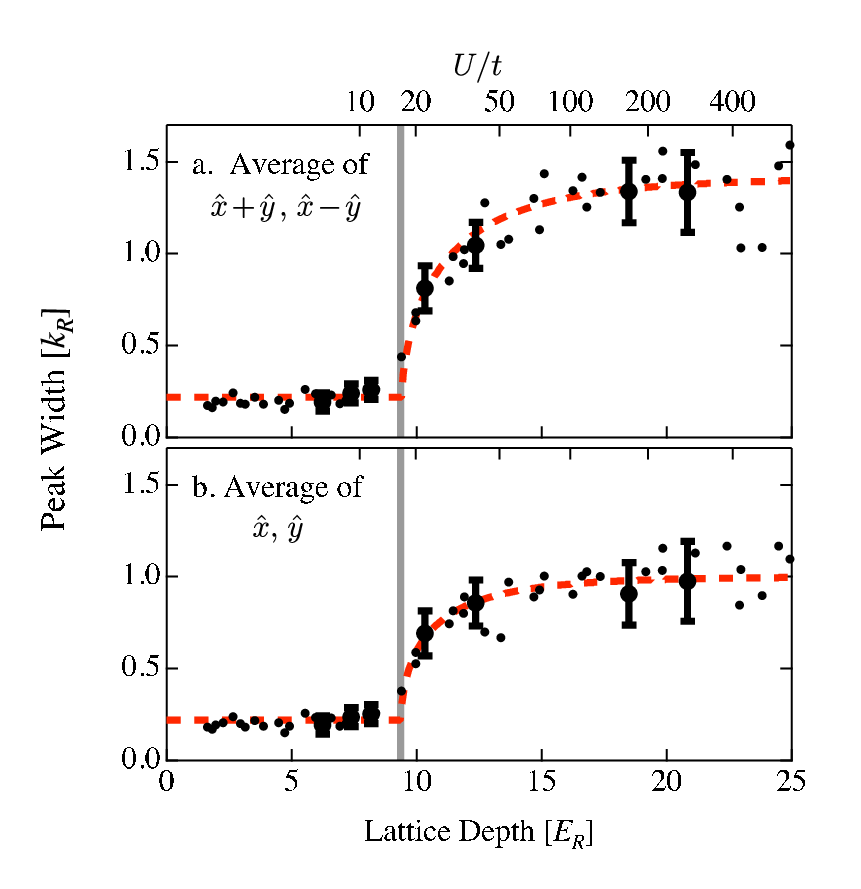}
\end{center}
\caption{Quasi-momentum width vs. $V$ (bottom axis) or $U/t$ (top axis).  The symbols denote the average FWHM of the quasi-momentum distribution along the axes of highest symmetry (Top: averaged along $\hat x$+$\hat y$ and $\hat x$-$\hat y$; Bottom: averaged along $\hat x$ and $\hat y$).  The small and large dots and uncertainties are as explained at Fig. \ref{fract}.  The red-dashed line is the horizontally displaced RPA momentum width, as discussed in the text, and the vertical grey line denotes the location of the SF-normal transition identified from the sudden increase in $\Gamma$.}
\label{width}
\end{figure}

A related characterization of the system is the FWHM $\Gamma$ of the full quasi-momentum distribution~\cite{Greiner2002,Wessel2004,Kato2007,PolletPrivate,Yi2007}.  Figure \ref{width} shows the width of the 2D distributions (see Ref.~\cite{Spielman2007}) as a function of $V$.  In the SF phase $\Gamma$ hardly depends on $V$ since the dominant feature of the distribution is the condensate peak.  $\Gamma$ only begins to change very close to the SF to normal phase transition when the heights of the two components of the bimodal distribution become comparable when the condensate disappears, consistent with calculations in homogenous and trapped systems~\cite{Wessel2004}.  We calculate $\Gamma$ in the MI phase from the RPA~\cite{Sengupta2005} quasi-momentum distribution which accurately describes the large $U/t$ limit ($\sqrt{2}\kr$ along $\hat x + \hat y$ and $\kr$ along $\hat x$).  In the RPA, $\Gamma$ is a function of $U/t$ in units of $\UoverTC$.  The red dashed lines are fits to the measured widths using two free parameters (joint between both panels): in the MI region we use the RPA functional form with $\UoverTC$ as the first fit parameter, and the constant width in the SF phase is the second.  We obtain $\UoverTC=16.7(20)$, in accord with the $\UoverTC=15.8(20)$ from our fit to the condensate fraction.

We identify the point when the condensate fraction vanishes (and $\Gamma$ abruptly increases) with the onset of the SF to normal transition, i.e., when a normal region begins to rapidly expand in our inhomogeneous system.  (Our measured visibility, computed as in Refs.~\cite{Gerbier2005,Gerbier2005a}, abruptly drops from near unity at $U/t\approx16$.)  Increasingly accurate numerical calculations give values of $\UoverTC$: $16.25(10)$~\cite{Wessel2004} and $16.77$~\cite{Kato2007}.  Perhaps most relevant are QMC calculations which include the effects of harmonic confinement; in this case, Wessel {\it et al}.~\cite{Wessel2004} find that a MI region first forms at $\UoverTC = 17.2$ (the exact value of $\UoverTC$ depends on the details of the harmonic potential).  Both values lie within our experimental uncertainty.

The calculations~\cite{Wessel2004, Kato2007} are at zero temperature, and while they agree with our observed $\UoverTC$, they do not predict a sudden increase in peak width or a vanishing condensate fraction at $\UoverTC$.  At $T=0$ and as $U/t$ increases past $\UoverTC$, where an inhomogeneous system first develops a unit-occupied Mott core, the shell of SF persists to large $U/t$.  Thus, at $T=0$, $f$ drops rapidly at $\UoverTC$, but does not vanish.  Our system, however, is at small but non-zero temperature, with a reduced condensate fraction of $\approx 45\%$ for small $V$.  Our MFT model shows that this temperatures quickly drives the SF shells to the normal phase as $U/t$ increases past $\UoverTC$.  As a result the SF shells rapidly disappear as the normal, and then Mott regions form (see Refs.~\cite{Rey2006,Ho2007,Gerbier2007}).  That this feature is seen in preliminary non-zero temperature QMC calculations~\cite{PolletPrivate} underscores the need for further non-zero temperature calculations to compare with experiment.  

This experiment constitutes the measurement of a single point of the non-zero temperature 2D BH phase diagram.  We expect future experiments will expand on this result at different temperatures, densities, in different dimensions, and in traps with more homogenous density distributions; new theory should aide in the interpretation of these experiments.

We appreciate enlightening conversations with C. A. R. Sa de Melo, N. Trivedi, L. Pollet, and  C. J. Williams.  We acknowledge the financial support of ODNI/IARPA, and ONR; and I.B.S. thanks the NIST/NRC program.


\begin{thebibliography}{27}
\expandafter\ifx\csname natexlab\endcsname\relax\def\natexlab#1{#1}\fi
\expandafter\ifx\csname bibnamefont\endcsname\relax
  \def\bibnamefont#1{#1}\fi
\expandafter\ifx\csname bibfnamefont\endcsname\relax
  \def\bibfnamefont#1{#1}\fi
\expandafter\ifx\csname citenamefont\endcsname\relax
  \def\citenamefont#1{#1}\fi
\expandafter\ifx\csname url\endcsname\relax
  \def\url#1{\texttt{#1}}\fi
\expandafter\ifx\csname urlprefix\endcsname\relax\def\urlprefix{URL }\fi
\providecommand{\bibinfo}[2]{#2}
\providecommand{\eprint}[2][]{\url{#2}}

\bibitem[{\citenamefont{Gerbier
  et~al.}(2005{\natexlab{a}})\citenamefont{Gerbier, Widera, F{\"o}lling,
  Mandel, Gericke, and Bloch}}]{Gerbier2005}
\bibinfo{author}{\bibfnamefont{F.}~\bibnamefont{Gerbier}},
  \bibinfo{author}{\bibfnamefont{A.}~\bibnamefont{Widera}},
  \bibinfo{author}{\bibfnamefont{S.}~\bibnamefont{F{\"o}lling}},
  \bibinfo{author}{\bibfnamefont{O.}~\bibnamefont{Mandel}},
  \bibinfo{author}{\bibfnamefont{T.}~\bibnamefont{Gericke}}, \bibnamefont{and}
  \bibinfo{author}{\bibfnamefont{I.}~\bibnamefont{Bloch}},
  \bibinfo{journal}{Physical Review Letters} \textbf{\bibinfo{volume}{95}},
  \bibinfo{pages}{050404} (\bibinfo{year}{2005}{\natexlab{a}}).

\bibitem[{\citenamefont{Spielman et~al.}(2007)\citenamefont{Spielman, Phillips,
  and Porto}}]{Spielman2007}
\bibinfo{author}{\bibfnamefont{I.~B.} \bibnamefont{Spielman}},
  \bibinfo{author}{\bibfnamefont{W.~D.} \bibnamefont{Phillips}},
  \bibnamefont{and} \bibinfo{author}{\bibfnamefont{J.~V.} \bibnamefont{Porto}},
  \bibinfo{journal}{Physical Review Letters} \textbf{\bibinfo{volume}{98}},
  \bibinfo{pages}{080404} (\bibinfo{year}{2007}).

\bibitem[{\citenamefont{Mun et~al.}(2007)\citenamefont{Mun, Medley, Campbell,
  Marcassa, Pritchard, and Ketterle}}]{Mun2007}
\bibinfo{author}{\bibfnamefont{J.}~\bibnamefont{Mun}},
  \bibinfo{author}{\bibfnamefont{P.}~\bibnamefont{Medley}},
  \bibinfo{author}{\bibfnamefont{G.~K.} \bibnamefont{Campbell}},
  \bibinfo{author}{\bibfnamefont{L.~G.} \bibnamefont{Marcassa}},
  \bibinfo{author}{\bibfnamefont{D.~E.} \bibnamefont{Pritchard}},
  \bibnamefont{and} \bibinfo{author}{\bibfnamefont{W.}~\bibnamefont{Ketterle}},
  \bibinfo{journal}{Physical Review Letters} \textbf{\bibinfo{volume}{99}},
  \bibinfo{eid}{150604} (\bibinfo{year}{2007}).

\bibitem[{\citenamefont{Jaksch et~al.}(1998)\citenamefont{Jaksch, Bruder,
  Cirac, Gardiner, and Zoller}}]{Jaksch1998}
\bibinfo{author}{\bibfnamefont{D.}~\bibnamefont{Jaksch}},
  \bibinfo{author}{\bibfnamefont{C.}~\bibnamefont{Bruder}},
  \bibinfo{author}{\bibfnamefont{J.~I.} \bibnamefont{Cirac}},
  \bibinfo{author}{\bibfnamefont{C.~W.} \bibnamefont{Gardiner}},
  \bibnamefont{and} \bibinfo{author}{\bibfnamefont{P.}~\bibnamefont{Zoller}},
  \bibinfo{journal}{Physical Review Letters} \textbf{\bibinfo{volume}{81}},
  \bibinfo{pages}{3108} (\bibinfo{year}{1998}).

\bibitem[{\citenamefont{K{\"o}hl et~al.}(2005)\citenamefont{K{\"o}hl, Moritz,
  St{\"o}ferle, Schori, and Esslinger}}]{Kohl2005}
\bibinfo{author}{\bibfnamefont{M.}~\bibnamefont{K{\"o}hl}},
  \bibinfo{author}{\bibfnamefont{H.}~\bibnamefont{Moritz}},
  \bibinfo{author}{\bibfnamefont{T.}~\bibnamefont{St{\"o}ferle}},
  \bibinfo{author}{\bibfnamefont{C.}~\bibnamefont{Schori}}, \bibnamefont{and}
  \bibinfo{author}{\bibfnamefont{T.}~\bibnamefont{Esslinger}},
  \bibinfo{journal}{Journal of Low Temperature Physics}
  \textbf{\bibinfo{volume}{138}}, \bibinfo{pages}{635} (\bibinfo{year}{2005}).

\bibitem[{\citenamefont{Greiner et~al.}(2002)\citenamefont{Greiner, Mandel,
  Esslinger, H{\"a}nsch, and Bloch}}]{Greiner2002}
\bibinfo{author}{\bibfnamefont{M.}~\bibnamefont{Greiner}},
  \bibinfo{author}{\bibfnamefont{O.}~\bibnamefont{Mandel}},
  \bibinfo{author}{\bibfnamefont{T.}~\bibnamefont{Esslinger}},
  \bibinfo{author}{\bibfnamefont{T.}~\bibnamefont{H{\"a}nsch}},
  \bibnamefont{and} \bibinfo{author}{\bibfnamefont{I.}~\bibnamefont{Bloch}},
  \bibinfo{journal}{Nature} \textbf{\bibinfo{volume}{415}}, \bibinfo{pages}{39}
  (\bibinfo{year}{2002}).

\bibitem[{\citenamefont{Gerbier
  et~al.}(2005{\natexlab{b}})\citenamefont{Gerbier, Widera, F{\"o}lling,
  Mandel, Gericke, and Bloch}}]{Gerbier2005a}
\bibinfo{author}{\bibfnamefont{F.}~\bibnamefont{Gerbier}},
  \bibinfo{author}{\bibfnamefont{A.}~\bibnamefont{Widera}},
  \bibinfo{author}{\bibfnamefont{S.}~\bibnamefont{F{\"o}lling}},
  \bibinfo{author}{\bibfnamefont{O.}~\bibnamefont{Mandel}},
  \bibinfo{author}{\bibfnamefont{T.}~\bibnamefont{Gericke}}, \bibnamefont{and}
  \bibinfo{author}{\bibfnamefont{I.}~\bibnamefont{Bloch}},
  \bibinfo{journal}{Physical Review A} \textbf{\bibinfo{volume}{72}},
  \bibinfo{pages}{053606} (\bibinfo{year}{2005}{\natexlab{b}}).

\bibitem[{\citenamefont{F{\"o}lling et~al.}(2006)\citenamefont{F{\"o}lling,
  Widera, M{\"u}ller, Gerbier, and Bloch}}]{Folling2006a}
\bibinfo{author}{\bibfnamefont{S.}~\bibnamefont{F{\"o}lling}},
  \bibinfo{author}{\bibfnamefont{A.}~\bibnamefont{Widera}},
  \bibinfo{author}{\bibfnamefont{T.}~\bibnamefont{M{\"u}ller}},
  \bibinfo{author}{\bibfnamefont{F.}~\bibnamefont{Gerbier}}, \bibnamefont{and}
  \bibinfo{author}{\bibfnamefont{I.}~\bibnamefont{Bloch}},
  \bibinfo{journal}{Physical Review Letters} \textbf{\bibinfo{volume}{97}},
  \bibinfo{pages}{060403} (\bibinfo{year}{2006}).

\bibitem[{\citenamefont{Wessel et~al.}(2004)\citenamefont{Wessel, Alet, Troyer,
  and Batrouni}}]{Wessel2004}
\bibinfo{author}{\bibfnamefont{S.}~\bibnamefont{Wessel}},
  \bibinfo{author}{\bibfnamefont{F.}~\bibnamefont{Alet}},
  \bibinfo{author}{\bibfnamefont{M.}~\bibnamefont{Troyer}}, \bibnamefont{and}
  \bibinfo{author}{\bibfnamefont{G.~G.} \bibnamefont{Batrouni}},
  \bibinfo{journal}{Physical Review A} \textbf{\bibinfo{volume}{70}},
  \bibinfo{pages}{053615} (\bibinfo{year}{2004}).

\bibitem[{\citenamefont{Campbell et~al.}(2006)\citenamefont{Campbell, Mun,
  Boyd, Medley, Leanhardt, Marcassa, Pritchard, and Ketterle}}]{Campbell2006}
\bibinfo{author}{\bibfnamefont{G.~K.} \bibnamefont{Campbell}},
  \bibinfo{author}{\bibfnamefont{J.}~\bibnamefont{Mun}},
  \bibinfo{author}{\bibfnamefont{M.}~\bibnamefont{Boyd}},
  \bibinfo{author}{\bibfnamefont{P.}~\bibnamefont{Medley}},
  \bibinfo{author}{\bibfnamefont{A.~E.} \bibnamefont{Leanhardt}},
  \bibinfo{author}{\bibfnamefont{L.}~\bibnamefont{Marcassa}},
  \bibinfo{author}{\bibfnamefont{D.~E.} \bibnamefont{Pritchard}},
  \bibnamefont{and} \bibinfo{author}{\bibfnamefont{W.}~\bibnamefont{Ketterle}},
  \bibinfo{journal}{Science} \textbf{\bibinfo{volume}{313}},
  \bibinfo{pages}{649} (\bibinfo{year}{2006}).

\bibitem[{\citenamefont{Hadzibabic et~al.}(2006)\citenamefont{Hadzibabic,
  Kr{\"u}ger, Cheneau, Battelier, and Dalibard}}]{Hadzibabic2006}
\bibinfo{author}{\bibfnamefont{Z.}~\bibnamefont{Hadzibabic}},
  \bibinfo{author}{\bibfnamefont{P.}~\bibnamefont{Kr{\"u}ger}},
  \bibinfo{author}{\bibfnamefont{M.}~\bibnamefont{Cheneau}},
  \bibinfo{author}{\bibfnamefont{B.}~\bibnamefont{Battelier}},
  \bibnamefont{and} \bibinfo{author}{\bibfnamefont{J.}~\bibnamefont{Dalibard}},
  \bibinfo{journal}{Nature} \textbf{\bibinfo{volume}{441}}
  (\bibinfo{year}{2006}).

\bibitem[{\citenamefont{Clad{\'e} and Helmerson}(2008)}]{Clade2008}
\bibinfo{author}{\bibfnamefont{P.}~\bibnamefont{Clad{\'e}}} \bibnamefont{and}
  \bibinfo{author}{\bibfnamefont{K.}~\bibnamefont{Helmerson}}
  (\bibinfo{year}{2008}), \bibinfo{note}{private communication}.

\bibitem[{\citenamefont{Kruger et~al.}(2007)\citenamefont{Kruger, Hadzibabic,
  and Dalibard}}]{Kruger2007}
\bibinfo{author}{\bibfnamefont{P.}~\bibnamefont{Kruger}},
  \bibinfo{author}{\bibfnamefont{Z.}~\bibnamefont{Hadzibabic}},
  \bibnamefont{and} \bibinfo{author}{\bibfnamefont{J.}~\bibnamefont{Dalibard}},
  \bibinfo{journal}{Physical Review Letters} \textbf{\bibinfo{volume}{99}},
  \bibinfo{eid}{040402} (\bibinfo{year}{2007}).

\bibitem[{\citenamefont{van Kempen et~al.}(2002)\citenamefont{van Kempen,
  Kokkelmans, Heinzen, and Verhaar}}]{vanKempen2002}
\bibinfo{author}{\bibfnamefont{E.~G.~M.} \bibnamefont{van Kempen}},
  \bibinfo{author}{\bibfnamefont{S.~J. J. M.~F.} \bibnamefont{Kokkelmans}},
  \bibinfo{author}{\bibfnamefont{D.~J.} \bibnamefont{Heinzen}},
  \bibnamefont{and} \bibinfo{author}{\bibfnamefont{B.~J.}
  \bibnamefont{Verhaar}}, \bibinfo{journal}{Phys. Rev. Lett.}
  \textbf{\bibinfo{volume}{88}}, \bibinfo{pages}{093201}
  (\bibinfo{year}{2002}).

\bibitem[{\citenamefont{Fisher et~al.}(1989)\citenamefont{Fisher, Weichman,
  Grinstein, and Fisher}}]{Fisher1989}
\bibinfo{author}{\bibfnamefont{M.~P.~A.} \bibnamefont{Fisher}},
  \bibinfo{author}{\bibfnamefont{P.~B.} \bibnamefont{Weichman}},
  \bibinfo{author}{\bibfnamefont{G.}~\bibnamefont{Grinstein}},
  \bibnamefont{and} \bibinfo{author}{\bibfnamefont{D.~S.}
  \bibnamefont{Fisher}}, \bibinfo{journal}{Physical Review B}
  \textbf{\bibinfo{volume}{40}}, \bibinfo{pages}{546} (\bibinfo{year}{1989}).

\bibitem[{\citenamefont{Krauth and Trivedi}(1991)}]{Krauth1991a}
\bibinfo{author}{\bibfnamefont{W.}~\bibnamefont{Krauth}} \bibnamefont{and}
  \bibinfo{author}{\bibfnamefont{N.}~\bibnamefont{Trivedi}},
  \bibinfo{journal}{Europhysics Letters} \textbf{\bibinfo{volume}{14}},
  \bibinfo{pages}{627} (\bibinfo{year}{1991}).

\bibitem[{\citenamefont{Elstner and Monien}(1999)}]{Elstner1999}
\bibinfo{author}{\bibfnamefont{N.}~\bibnamefont{Elstner}} \bibnamefont{and}
  \bibinfo{author}{\bibfnamefont{H.}~\bibnamefont{Monien}},
  \bibinfo{journal}{arXiv:cond-mat} p. \bibinfo{pages}{9905367}
  (\bibinfo{year}{1999}).

\bibitem[{\citenamefont{Kato et~al.}(2007)\citenamefont{Kato, Kawashima, and
  Trivedi}}]{Kato2007}
\bibinfo{author}{\bibfnamefont{Y.}~\bibnamefont{Kato}},
  \bibinfo{author}{\bibfnamefont{N.}~\bibnamefont{Kawashima}},
  \bibnamefont{and} \bibinfo{author}{\bibfnamefont{N.}~\bibnamefont{Trivedi}}
  (\bibinfo{year}{2007}).

\bibitem[{\citenamefont{Yi et~al.}(2007)\citenamefont{Yi, Lin, and
  Duan}}]{Yi2007}
\bibinfo{author}{\bibfnamefont{W.}~\bibnamefont{Yi}},
  \bibinfo{author}{\bibfnamefont{G.-D.} \bibnamefont{Lin}}, \bibnamefont{and}
  \bibinfo{author}{\bibfnamefont{L.-M.} \bibnamefont{Duan}},
  \bibinfo{journal}{arXiv} p. \bibinfo{pages}{0705.4352v1}
  (\bibinfo{year}{2007}).

\bibitem[{\citenamefont{Diener et~al.}(2007)\citenamefont{Diener, Zhou, Zhai,
  and Ho}}]{Diener2007}
\bibinfo{author}{\bibfnamefont{R.~B.} \bibnamefont{Diener}},
  \bibinfo{author}{\bibfnamefont{Q.}~\bibnamefont{Zhou}},
  \bibinfo{author}{\bibfnamefont{H.}~\bibnamefont{Zhai}}, \bibnamefont{and}
  \bibinfo{author}{\bibfnamefont{T.-L.} \bibnamefont{Ho}},
  \bibinfo{journal}{Physical Review Letters} \textbf{\bibinfo{volume}{98}},
  \bibinfo{pages}{180404} (\bibinfo{year}{2007}).

\bibitem[{\citenamefont{Gerbier}(2007)}]{Gerbier2007}
\bibinfo{author}{\bibfnamefont{F.}~\bibnamefont{Gerbier}},
  \bibinfo{journal}{Physical Review Letters} \textbf{\bibinfo{volume}{99}},
  \bibinfo{pages}{120405} (\bibinfo{year}{2007}).

\bibitem[{\citenamefont{Ho and Zhou}(2007)}]{Ho2007}
\bibinfo{author}{\bibfnamefont{T.-L.} \bibnamefont{Ho}} \bibnamefont{and}
  \bibinfo{author}{\bibfnamefont{Q.}~\bibnamefont{Zhou}},
  \bibinfo{journal}{Physical Review Letters} \textbf{\bibinfo{volume}{99}},
  \bibinfo{pages}{120404} (\bibinfo{year}{2007}).

\bibitem[{\citenamefont{Sebby-Strabley
  et~al.}(2006)\citenamefont{Sebby-Strabley, Anderlini, Jessen, and
  Porto}}]{Sebby-Strabley2006}
\bibinfo{author}{\bibfnamefont{J.}~\bibnamefont{Sebby-Strabley}},
  \bibinfo{author}{\bibfnamefont{M.}~\bibnamefont{Anderlini}},
  \bibinfo{author}{\bibfnamefont{P.~S.} \bibnamefont{Jessen}},
  \bibnamefont{and} \bibinfo{author}{\bibfnamefont{J.~V.} \bibnamefont{Porto}},
  \bibinfo{journal}{Phys. Rev. A} \textbf{\bibinfo{volume}{73}},
  \bibinfo{pages}{033605} (\bibinfo{year}{2006}).

\bibitem[{\citenamefont{Ovchinnikov et~al.}(1998)\citenamefont{Ovchinnikov,
  M\"uller, Doery, Vredenbregt, Helmerson, Rolston, and
  Phillips}}]{Ovchinnikov1998}
\bibinfo{author}{\bibfnamefont{Y.~B.} \bibnamefont{Ovchinnikov}},
  \bibinfo{author}{\bibfnamefont{J.~H.} \bibnamefont{M\"uller}},
  \bibinfo{author}{\bibfnamefont{M.~R.} \bibnamefont{Doery}},
  \bibinfo{author}{\bibfnamefont{E.~J.~D.} \bibnamefont{Vredenbregt}},
  \bibinfo{author}{\bibfnamefont{K.}~\bibnamefont{Helmerson}},
  \bibinfo{author}{\bibfnamefont{S.~L.} \bibnamefont{Rolston}},
  \bibnamefont{and} \bibinfo{author}{\bibfnamefont{W.~D.}
  \bibnamefont{Phillips}}, \bibinfo{journal}{Phys. Rev. Lett.}
  \textbf{\bibinfo{volume}{83}}, \bibinfo{pages}{284} (\bibinfo{year}{1998}).

\bibitem[{\citenamefont{Pollet and Troyer}()}]{PolletPrivate}
\bibinfo{author}{\bibfnamefont{L.}~\bibnamefont{Pollet}} \bibnamefont{and}
  \bibinfo{author}{\bibfnamefont{M.}~\bibnamefont{Troyer}},
  \bibinfo{note}{private communication}.

\bibitem[{\citenamefont{Sengupta and Dupuis}(2005)}]{Sengupta2005}
\bibinfo{author}{\bibfnamefont{K.}~\bibnamefont{Sengupta}} \bibnamefont{and}
  \bibinfo{author}{\bibfnamefont{N.}~\bibnamefont{Dupuis}},
  \bibinfo{journal}{Physical Review A} \textbf{\bibinfo{volume}{71}},
  \bibinfo{pages}{033629} (\bibinfo{year}{2005}).

\bibitem[{\citenamefont{Sheshadri et~al.}(1993)\citenamefont{Sheshadri,
  Krishnamurthy, Pandit, and Ramakrishnan}}]{Sheshadri1993}
\bibinfo{author}{\bibfnamefont{K.}~\bibnamefont{Sheshadri}},
  \bibinfo{author}{\bibfnamefont{H.~R.} \bibnamefont{Krishnamurthy}},
  \bibinfo{author}{\bibfnamefont{R.}~\bibnamefont{Pandit}}, \bibnamefont{and}
  \bibinfo{author}{\bibfnamefont{T.~V.} \bibnamefont{Ramakrishnan}},
  \bibinfo{journal}{Europhysics Letters} \textbf{\bibinfo{volume}{22}},
  \bibinfo{pages}{257} (\bibinfo{year}{1993}).

\bibitem[{\citenamefont{Rey et~al.}(2006)\citenamefont{Rey, Pupillo, and
  Porto}}]{Rey2006}
\bibinfo{author}{\bibfnamefont{A.~M.} \bibnamefont{Rey}},
  \bibinfo{author}{\bibfnamefont{G.}~\bibnamefont{Pupillo}}, \bibnamefont{and}
  \bibinfo{author}{\bibfnamefont{J.~V.} \bibnamefont{Porto}},
  \bibinfo{journal}{Physical Review A} \textbf{\bibinfo{volume}{73}},
  \bibinfo{eid}{023608} (\bibinfo{year}{2006}).

\end{thebibliography}
\end{document}